\documentclass[aps,prd,twocolumn,groupedaddress,nofootinbib,showpacs,eqsecnum]{revtex4}

\usepackage{euscript,graphicx,amssymb,subfigure}

\usepackage{amsfonts}
\usepackage{amsmath}
\usepackage{amssymb}
\usepackage{graphicx}
\usepackage{euscript}
\usepackage{color}
\usepackage{subfigure}

\newcommand{\goodgap}{\hspace{\subfigtopskip} \hspace{\subfigbottomskip}}

\begin{document}

\title{Cosmography in $f(T)$-gravity}

\author{S. Capozziello$^{1,2}$\footnote{{\tt capozziello@na.infn.it}}, V. F. Cardone$^{3}$\footnote{{\tt winnyenodrac@gmail.com}}
H. Farajollahi$^{4}$
\footnote{{\tt hosseinf@guilan.ac.ir}},
A. Ravanpak$^{4}$
\footnote{{\tt aravanpak@guilan.ac.ir}} }

\affiliation{$^1$ Dipartimento di Scienze Fisiche, Universit\`{a} di Napoli ``Federico II'', Compl. Univ. Monte S. Angelo, Ed.N, Via Cinthia, I-80126 Napoli, Italy}

\affiliation{$^2$ I.N.F.N. - Sez. di Napoli, Compl. Univ. Monte S. Angelo, Ed.G, Via Cinthia, I-80126 Napoli, Italy}

\affiliation{$^3$ I.N.A.F. - Osservatorio Astronomico di Roma, via Frascati 33, 00040\,-\,Monte Porzio Catone (Roma), Italy}

\affiliation{$^4$Department of Physics, University of Guilan, Rasht, Iran.}

\begin{abstract}

Being based on the only assumption that the universe is homogenous and isotropic on large scales, cosmography is an ideal tool to investigate the cosmic expansion history in a almost model-independent way. Fitting the data on the luminosity distance and Baryon Acoustic Oscillations allows to determine the confidence ranges for the cosmographic parameters hence giving some quantitative constraints that a whatever theory has to fulfill. As an application, we consider here the case of teleparallel gravity (TEGR) also referred to as $f(T)$ gravity. To this end, we first work out analytical expressions to express the present day values of $f(T)$ derivatives as a function of the cosmographic parameters which hold under quite general and physically motivated conditions. We then use the constraints coming from cosmography to find out the confidence ranges for $f(T)$ derivatives up to the fifth order and show how these can be used to check the viability of  given TEGR models without the need to explicitly solve the second order dynamic equations.

\end{abstract}


\date{\today}
\maketitle

\section{Introduction}

Various cosmological observations, including the Type Ia Supernova (Sne Ia) \cite{Riess}, the cosmic microwave background radiation \cite{Spergel} and the large scale structure \cite{Tegmark,Eisenstein}, et al., have revealed that the universe is undergoing an accelerating expansion and it entered this accelerating phase only in the near past. This unexpected observed phenomenon poses one of the most puzzling problems in cosmology today. Usually, it is assumed that there exists dark energy (DE) in our universe, as an exotic energy component with negative pressure which dominates the universe filled with cold dark matter (CDM) and drives the universe to an accelerating expansion at recent times.

The simplest and most appealing candidate for DE is the vacuum energy (cosmological constant, $\Lambda$) with a constant equation of state (EoS) parameter, $-1$. This model is in general agreement with the current astronomical observations, but has difficulties to reconcile the small observational value of DE density to that coming from quantum field theories. This is called the cosmological constant problem \cite{Carroll}. Recently it was shown that $\Lambda$CDM model may also suffer from an age problem \cite{Yang}. It is thus natural to pursue alternative
possibilities to explain the current acceleration of the universe. Observing the small deviations for the EoS parameter from $-1$ requires a description of the DE that allows the EoS to evolve across the phantom divide line $-1$ possibly multiple times. The present data seem to slightly favor an evolving DE with the EoS parameter crossing $-1$ from above to below in the near past \cite{Feng}. One may take the observed accelerating expansion as a signal of the breakdown of our understanding to the laws of gravitation and, thus, a modification of the gravity theory is needed.

Over the past decade numerous DE models have been proposed, such as quintessence \cite{Caldwell}, phantom \cite{Caldwell2}, k-essence \cite{Armendariz}, tachyon \cite{Padmanabhan}, quintom \cite{Feng, eli}; as well as the Chaplygin gas \cite{Kamenshchik} and the generalized Chaplygin gas (GCG) \cite{Bento}, the holographic DE \cite{Cohen}, the new agegraphic DE \cite{Wei}, the Ricci DE \cite{Gao} and so on.

In addition, the extensions to gravity by making the action a function of the spacetime curvature scalar $R$, $f(R)$ \cite{capfra,Nojiri,book,sotfa},
or other curvature invariants \cite{Mota}, by coupling the Ricci scalar to a scalar field \cite{Farajollahi}, by introducing a vector field contribution \cite{Zuntz}, or by using properties of gravity in higher dimensional spacetimes have widely been investigated \cite{Camera}. Among  $f(R)$ models, there are models that are verified by all the observational and theoretical constraints and exhibit universe acceleration and phantom crossing \cite{noj1}--\cite{noj5}.

In a different approach, avoiding the curvature defined via the Levi-Civita connection, one could explore an alternative way and use the Weitzenb\"{o}ck connection that has no curvature but instead torsion. This has the property that the torsion is formed completely from products of first derivatives of the tetrad, with no second derivatives appearing in the torsion tensor. This approach was considered originally by Einstein in 1928 \cite{Einstein,Einstein2}, as "Teleparallelism". It is closely related to standard General Relativity, differing only in "boundary terms" involving total derivatives in the action. The theory is naturally formulated by gauging external (spacetime) translation and underline the Weitzenb\"{o}ck spacetime characterized by the metricity condition and by the vanishing of the curvature tensor. Translations are closely related to the group of general coordinate transformations which underlies General Relativity. The theory possesses a number of attractive features both from the geometrical and physical viewpoints \cite{Hoff}--\cite{Yang2}.

Some models based on modification of the teleparallel equivalent of General Relativity (TEGR) are presented as an alternative to inflationary models without an inflaton \cite{Ferraro2,Poplawski} and DE models for the acceleration of the universe \cite{Wu}--\cite{Yang2} where dark torsion (DT) is responsible for the observed acceleration of the universe, and the field equations are always $2nd$ order equations. This property makes these theories simpler than the dynamical equations resulting in $f(R)$ gravity among other advantages. It has been shown that in theories of generalised TEGR, whose Lagrangians are algebraic
functions of the usual teleparallel Lagrangian, the action and the field equations are not invariant
under local Lorentz transformations \cite{lorentz}. The authors also argue that the usual teleparallel Lagrangian equivalent to General Relativity, is just a special case.

It is worth noticing that all the above models such as dark energy, $f(R)$-gravity and $f(T)$-gravity have shown to be, in broad sense,  in agreement with observational data. As a consequence, unless higher precision probes of the expansion rate and the growth of structure will be available, these  rival approaches could not be discriminated. This degeneration about the theoretical background suggests that a more conservative approach to the problem of cosmic acceleration, relying on as less model dependent quantities as possible, is welcome.

A possible solution could be to come back to the cosmography rather than finding out solutions of the Friedmann equations and testing them. Being only related to the derivatives of the scale factor, the cosmographic parameters make it possible to fit the data on the distance - redshift relation without any a priori assumption on the underlying cosmological model. In this case, the only assumption is that the metric is the Robertson - Walker one. Almost a century after Hubble discovery of the expansion of the universe, we could now extend cosmography
beyond the search for the value of the Hubble constant. The SNeIa Hubble diagram extends up to $z = 1.7$ thus invoking the need for, at least, a fifth order Taylor expansion of the scale factor in order to give a reliable approximation of the distance - redshift relation. As a consequence, it could be, in principle, possible to estimate up to five cosmographic parameters, although the still too small dataset available does not allow to get a precise and realistic determination of all of them.

Once these quantities have been determined, one could use them to put constraints on the models. In a sense, we are reverting the usual approach consisting in deriving the cosmographic parameters as a sort of byproduct of an assumed theory. Here, we follow the other way around expressing the model characterizing quantities as a function of the cosmographic parameters. Such a program has been particularly suited for the study of fourth order theories of gravity, i.e. $f(R)$-gravity \cite{Capozziello:2008qc,blcc10}. As it is well known, the mathematical difficulties entering the solution of fourth order field equations make it quite problematic to find out analytical expressions for the scale factor and hence predict the values of the cosmographic parameters. But, nobody has still studied this procedure in $f(T)$ gravity. A key role in both $f(R)$ and $f(T)$ scenarios is played by the choice of the $f(R)$ or the $f(T)$ function. Under quite general hypotheses, we will derive useful relations among the cosmographic parameters and the present day value of $f^{(n)}(T) = d^nf/dT^n$, with $n = 0,...,5$, whatever the analytic form of $f(T)$ is. These relations will then allow to constrain the $f(T)$ derivatives, provided model independent constraints on the cosmographic parameters are available.

The layout of the paper is as follows. In Section II we will review the $f(T)$ gravity. In Section III we will introduce the basic notions of the cosmographic parameters. Section IV contains the main result of the paper demonstrating how the $f(T)$ derivatives can be related to the cosmographic parameters, while, in Section V, we use these relations and previous constraints on the cosmographic parameters to derive model independent estimates of the present day values of the $f(T)$ derivatives. As a further application, Section VI shows how these latter constraints can be used to observationally validate a given class of TEGR models without the need to solve the field equations. We summarize and conclude in Section VII.

\section{$f(T)$ gravity}

Teleparallelism uses as dynamical object a vierbein
field $e_i(x^\mu)$, $i = 0, 1, 2, 3$, which is an orthonormal
basis for the tangent space at each point $x^\mu$ of the manifold: $e_i . e_j=\eta_{ij}$, where $\eta_{ij}=diag(1,-1,-1,-1)$. Each vector $e_i$ can be described by its components $e^\mu_i$, $\mu=0,1,2,3$ in a coordinate basis; i.e. $e_i=e^\mu_i\partial_\mu$.
Notice that latin indexes refer to the tangent space,
while greek indexes label coordinates on the manifold.
The metric tensor is obtained from the dual vierbein as $g_{\mu\nu}(x)=\eta_{ij} e^i_\mu(x)e^j_\nu(x)$. Differing from General Relativity,
which uses the torsionless Levi-Civita connection,
Teleparallelism uses the curvatureless Weitzenb\"{o}ck connection, whose non-null torsion is
\begin{equation}\label{torsion}
    T^\lambda_{\mu\nu}=\hat{\Gamma}^\lambda_{\nu\mu}-\hat{\Gamma}^\lambda_{\mu\nu}=e^\lambda_i(\partial_\mu e^i_\nu - \partial_\nu e^i_\mu).
\end{equation}
This tensor encompasses all the information about the
gravitational field. The TEGR Lagrangian is built with
the torsion (\ref{torsion}) and its dynamical equations for the vierbein
imply the Einstein equations for the metric. The
teleparallel Lagrangian is
\begin{equation}\label{lagrangian}
    T={S_\rho}^{\mu\nu}{T^\rho}_{\mu\nu},
\end{equation}
where
\begin{equation}\label{s}
    {S_\rho}^{\mu\nu}=\frac{1}{2}({K^{\mu\nu}}_\rho+\delta^\mu_\rho {T^{\theta\nu}}_\theta-\delta^\nu_\rho {T^{\theta\mu}}_\theta)
\end{equation}
and ${K^{\mu\nu}}_\rho$ is the contorsion tensor
\begin{equation}\label{contorsion}
    {K^{\mu\nu}}_\rho=-\frac{1}{2}({T^{\mu\nu}}_\rho-{T^{\nu\mu}}_\rho-{T_\rho}^{\mu\nu}),
\end{equation}
which equals the difference between Weitzenb\"{o}ck and Levi-Civita connections.

In this work the gravitational field will be driven by a Lagrangian density that is a function of $T$. Thus the action reads
\begin{equation}\label{action}
    I = \frac{1}{16\pi G}\int{d^4xef(T)},
\end{equation}
where $e=det(e^i_\mu)=\sqrt{-g}$. The case $f(T)=T$ corresponds to TEGR. If matter couples to the metric in the standard form then the variation of the action with respect to the vierbein leads to the equations \cite{Ferraro}
\begin{eqnarray}
  && e^{-1}\partial_\mu(e{S_i}^{\mu\nu})f'(T)-e_i^\lambda {T^\rho}_{\mu\lambda}{S_\rho}^{\nu\mu}f'(T) \nonumber \\ & & +{S_i}^{\mu\nu}\partial_\mu(T)f''(T) +\frac{1}{4}e^\nu_if(T)=4\pi G{e_i}^\rho {T_\rho}^\nu, \label{equations}
\end{eqnarray}
where a prime denotes differentiation with respect to $T$,
${S_i}^{\mu\nu}={e_i}^\rho {S_\rho}^{\mu\nu}$ and $T_{\mu\nu}$ is the matter energy-momentum
tensor.

We will assume a flat homogeneous and isotropic FRW universe, so
\begin{equation}\label{metric}
    e^i_\mu = diag(1, a(t), a(t), a(t)),
\end{equation}
where $a(t)$ is the cosmological scale factor. By using (\ref{torsion}), (\ref{lagrangian}), (\ref{s}) and (\ref{contorsion}) we obtain
\begin{equation}\label{lt}
    T=-6H^2,
\end{equation}
where $H=\frac{\dot a}{a}$ is the Hubble parameter. The substitution of the vierbein (\ref{metric}) in (\ref{equations})
for $i=0=\nu$ yields
\begin{equation}\label{friedmann}
    12H^2f'(T)+f(T)=16\pi G\rho.
\end{equation}
Besides, the equation $i=1=\nu$ is
\begin{equation}\label{acceleration}
    48H^2f''(T)\dot{H}-f'(T)[12H^2+4\dot{H}]-f(T)=16\pi Gp.
\end{equation}
In Eqs. (\ref{friedmann}) and (\ref{acceleration}), $\rho$ and $p$ are the dark matter energy density and pressure, respectively. It can be easily derived that they accomplish the conservation equation
\begin{equation}\label{conservation}
    \dot{\rho}+3H(\rho+p)=0.
\end{equation}
We can rewrite Eqs. (\ref{friedmann}) and (\ref{acceleration}) as the usual form
\begin{equation}\label{modfri}
    H^2=\frac{8\pi G}{3}(\rho+\rho_T),
\end{equation}
\begin{equation}\label{modacce}
    2\dot H+3H^2=-\frac{8\pi G}{3}(p+p_T)
\end{equation}
where
\begin{equation}\label{rhoT}
    \rho_T=\frac{1}{16\pi G}[2Tf'(T)-f(T)-T/2],
\end{equation}
\begin{equation}\label{pT}
    p_T=\frac{1}{16\pi G}[2\dot H(4Tf''(T)+2f'(T)-1)]-\rho_T.
\end{equation}
are the torsion contributions to the energy density and pressure. Then, by using Eqs. (\ref{rhoT}) and (\ref{pT}), we can define
the effective torsion equation of state as
\begin{equation}\label{omegaeff}
    \omega_{T}\equiv\frac{p_T}{\rho_T}=-1+\frac{4\dot H(4Tf''(T)+2f'(T)-1)}{4Tf'(T)-2f(T)-T}.
\end{equation}
This could be, in principle, related to the observed acceleration of the universe.

\section{cosmographic parameters}

Standard candles (such as SNeIa and, to a limited extent, gamma ray bursts) are ideal tools in modern cosmology since they make it possible to reconstruct the Hubble diagram, i.e. the redshift - distance relation up to high redshift values. It is then customary to assume a parameterized model (such as the concordance $\Lambda$CDM one, or any other kind of dark energy scenario) and contrasting it against the data to check its viability and constraints its characterizing parameters. As it is clear, such an approach is model dependent so that some doubts always remain on the validity of the constraints on derived quantities as the present day values of the deceleration parameter and the age of the universe. In order to overcome such a problem, one may resort to cosmography, i.e. expanding the scale factor in Taylor series with respect to the cosmic time. Such an expansions leads to a distance - redshift relation which only relies on the assumption of the Robertson -Walker metric thus being fully model independent since it does not depend on the particular form of the solution of cosmic equations. To this aim, it is convenient to introduce the following functions:
\begin{eqnarray}\label{par}
  H &=& \frac{1}{a}\frac{da}{dt} \\
  q &=& -\frac{1}{a}\frac{d^2a}{dt^2}H^{-2} \\
  j &=& \frac{1}{a}\frac{d^3a}{dt^3}H^{-3} \\
  s &=& \frac{1}{a}\frac{d^4a}{dt^4}H^{-4} \\
  l &=& \frac{1}{a}\frac{d^5a}{dt^5}H^{-5}
\end{eqnarray}
which are usually referred to as the Hubble, deceleration, jerk, snap and lerk parameters, respectively. Their present day values (which we will denote with a subscript 0) may be used to characterize the evolutionary status of the universe. For instance, $q_0 < 0$ denotes an accelerated expansion, while $j_0$ allows to discriminate among different accelerating models.

It is then a matter of algebra to demonstrate the following useful relations:
\begin{equation}\label{dotH}
    \dot H=-H^2(1+q),
\end{equation}
\begin{equation}\label{ddotH}
    \ddot H=H^3(j+3q+2),
\end{equation}
\begin{equation}\label{dddotH}
    \dddot H=H^4[s-4j-3q(q+4)-6],
\end{equation}
\begin{equation}\label{ddddotH}
    H^{(iv)}=H^5[l-5s+10(q+2)j+30(q+2)q+24],
\end{equation}
where a dot denotes derivative with respect to the cosmic time $t$ and $H^{(iv)}=d^4H/dt^4$. Eqs.(\ref{dotH}) - (\ref{ddddotH}) make it possible to relate the derivative of the Hubble parameter to the other cosmographic parameters.

\section{$f(T)$ derivatives vs cosmography}

Rather than choosing a parameterized expression for $f(T)$ and then numerically solving modified Friedmann equations for given values of the boundary conditions, we try to relate the present day values of its derivatives to the cosmographic parameters $(q_0, j_0, s_0, l_0)$ so that constraining them in a model independent way gives us a hint for what kind of $f(T)$ model could be able to fit the observed Hubble diagram.

As a preliminary step, it is worth considering Eq. (\ref{lt}). Differentiating with respect to $t$, we easily get the following relations:
\begin{eqnarray}
  \dot T &=& -12H\dot H, \label{dif1T} \\
  \ddot T &=& -12[\dot H^2+H\ddot H], \\ \label{dif2T}
  \dddot T &=& -12[3\dot H\ddot H+H\dddot H], \\ \label{dif3T}
  T^{(iv)} &=& -12[3\ddot H^2+4\dot H\dddot H+HH^{(iv)}] \label{dif4T}
\end{eqnarray}

The modified Friedmann Eqs. (\ref{friedmann}) and (\ref{acceleration}) can be rewritten as
\begin{equation}\label{friedfried}
    H^2=\frac{-1}{12f'(T)}[T\Omega_{m}+f(T)]
\end{equation}
and
\begin{equation}\label{acceacce}
    \dot H=\frac{1}{4f'(T)}[T\Omega_{m}-4H\dot{T}f''(T)]
\end{equation}
where dot denotes derivative with respect to the cosmic time $t$ and $\Omega_m$ represents the dimensionless matter density parameter. However, in order to enter other cosmographic parameters we have to differentiate from Eq.(\ref{acceacce}) three more times. We thus get:

\begin{widetext}

\begin{equation}\label{expddH}
    \ddot H=\frac{\Omega_m}{4Hf'(T)}[H\dot T-T(3H^2+2\dot H)]-\frac{1}{f'(T)}[(2\dot H\dot T+H\ddot T)f''(T)+H\dot T^2f'''(T)],
\end{equation}
\begin{eqnarray}
   \dddot H&=&\frac{\Omega_m}{4H^2f'(T)}[T(9H^4+6H^2\dot H+4\dot H^2)-H\dot T(3\dot H+6H^2)+H(H\ddot T-2\ddot H T)] \nonumber \\ &-&\frac{1}{Hf'(T)}[\dot H\ddot Hf'(T)+(2\dot H^2\dot T+3H\ddot H\dot T+4H\dot H\ddot T+H^2\dddot T)f''(T)+H^2\dot T^3f^{(iv)}(T) \nonumber \\&+&H\dot T(4\dot H\dot T+3H\ddot T)f'''(T)], \label{expdddH}
\end{eqnarray}
and
\begin{eqnarray}
  H^{(iv)}&=&\frac{\Omega_m}{4H^3f'(T)}[T(10H\dot H\ddot H+12H^3\ddot H-27H^6-12H^2\dot H^2-8\dot H^3-2H^2\dddot H)+H^3\dddot T \nonumber \\ &+&H^2\dot T(9H\dot H+27H^3-5\ddot H)-3H^2\ddot T(3H^2+\dot H)+7H\dot H^2\dot T]\nonumber \\ &-&\frac{1}{H^2f'(T)}[(3H\dot H\dddot H+\dot H^2\ddot H+H\ddot H^2)f'(T)+H^2\dot T^2(7\dot H\dot T+6H\ddot T)f^{(iv)}(T) \nonumber \\ &+&(4H^2\dddot H\dot T+2\dot H^3\dot T+7H^2\dot H\dddot T+10H\dot H^2\ddot T+7H^2\ddot H\ddot T+11H\dot H\ddot H\dot T+H^3T^{(iv)})f''(T) \nonumber \\ &+&H(10\dot H^2\dot T^2+7H\ddot H\dot T^2+21H\dot H\dot T\ddot T+3H^2\ddot T^2+4H^2\dot T\dddot T)f'''(T) \nonumber \\ &+&H^3\dot T^4f^{(v)}(T)], \label{expddddH}
\end{eqnarray}
with $f^{(iv)}(T)=d^4f(T)/dT^4$ and $f^{(v)}(T)=d^5f(T)/dT^5$.

\end{widetext}

Since the last five equations have to hold along the full evolutionary history of the universe, they naively hold also at the present day. As a consequence, we may evaluate them in $t = t_0$ thus easily obtaining:
\begin{equation}\label{friedcosmo}
    H_0^2=\frac{-1}{12f'(T_0)}[T_0\Omega_{m0}+f(T_0)]
\end{equation}
\begin{equation}\label{accecosmo}
    \dot{H_0}=\frac{1}{4f'(T_0)}[T_0\Omega_{m0}-4H_0\dot{T_0}f''(T_0)].
\end{equation}
and so on for the next three ones.

So, we have five equations, i.e. (\ref{friedcosmo}), (\ref{accecosmo}) and (\ref{expddH})-(\ref{expddddH}) evaluated at the present day. We call these, ``final equations'' which will turn out to be useful in the following. But, one further relation is needed in order to close the system and determine the six unknown quantities $f(T_0)$, $f'(T_0)$, $f''(T_0)$, $f'''(T_0)$, $f^{(iv)}(T_0)$ and $f^{(v)}(T_0)$. This may be easily obtained by noticing that, inserting back the physical units, equation (\ref{friedmann}) reads:
\begin{equation}\label{backfried}
    H^2=\frac{8\pi G}{6f'(T)}\left(\rho-\frac{f(T)}{16\pi G}\right),
\end{equation}
which clearly shows that, in $f(T)$ gravity, the Newtonian gravitational constant $G_N$ has to be replaced by an effective (time dependent) coupling $G_{eff}$. However, the present day value of Newtonian gravitational constant has to be recovered, and then:
\begin{equation}\label{constraint}
    G_{eff}(z=0)=G_N\rightarrow f'(T_0)=1\,, 
\end{equation}
which means the recovery of TEGR. In other words, the choice $f(T)=T$ gives rise to $T=-6H^2$ (see Eq.(\ref{lt})), and then Eq.(\ref{backfried}) reduces to the standard
\begin{equation}
H^2=\frac{8\pi G}{3}\rho\,.
\end{equation}

Let us now suppose that $f(T)$ may be well approximated by its fifth order Taylor expansion in $T-T_0$, i.e. we set:
\begin{eqnarray}
    f(T)& = &f(T_0)+f'(T_0)(T-T_0)+\frac{1}{2}f''(T_0)(T-T_0)^2 \nonumber \\
    & + & \frac{1}{6}f'''(T_0)(T-T_0)^3 + \frac{1}{24}f^{(iv)}(T_0)(T-T_0)^4 \nonumber \\
    & + & \frac{1}{120}f^{(v)}(T_0)(T-T_0)^5 . \label{taylor}
\end{eqnarray}
In such an approximation, it is $f^{(n)}(T)=d^nf/dT^n=0$ for $n\geq6$.

Evaluating Eqs.(\ref{dif1T}) - (\ref{dif4T}) and (\ref{lt}) at the present time and using Eqs.(\ref{dotH}) - (\ref{ddddotH}), one gets:
\begin{equation}\label{T}
    T_0=-6H_0^2,
\end{equation}
\begin{equation}\label{dotT}
  \dot{T_0}=12H_0^3(1+q_0),
\end{equation}
\begin{equation}\label{ddotT}
  \ddot{T_0}=-12H_0^4[q_0(q_0+5)+j_0+3],
\end{equation}
\begin{equation}\label{dddotT}
  \dddot{T_0}=-12H_0^5[s_0-j_0(3q_0+7)-3q_0(4q_0+9)-12],
\end{equation}
\begin{eqnarray}\label{ddddotT}
  T^{(iv)}_0 & = & -12H_0^6[l_0-s_0(4q_0+9)+j_0(3j_0+44q_0+48) \nonumber \\
  & + & 3q_0(4q_0^2+39q_0+56)+60].
\end{eqnarray}
After inserting all of these into the ``final equations'', we can solve them under the constraint (\ref{constraint}) with respect to the present day values of $f(T)$ and its derivatives up to the fifth order. After some algebra, one ends up with the desired result:

\begin{widetext}

\begin{equation}\label{fT}
    \frac{f(T_0)}{6H_0^2}=\Omega_{m0}-2,
\end{equation}
\begin{equation}\label{f1T}
    f'(T_0)=1,
\end{equation}
\begin{equation}\label{f2T}
    \frac{f''(T_0)}{(6H_0^2)^{-1}}=\frac{-3\Omega_{m0}}{4(1+q_0)}+\frac{1}{2},
\end{equation}
\begin{equation}\label{f3T}
    \frac{f'''(T_0)}{(6H_0^2)^{-2}}=\frac{-3\Omega_{m0}(3q_0^2+6q_0+j_0+2)}{8(1+q_0)^3}+\frac{3}{4},
\end{equation}
\begin{eqnarray}\label{f4T}
    \frac{f^{(iv)}(T_0)}{(6H_0^2)^{-3}}&=&\frac{-3\Omega_{m0}}{16(1+q_0)^5}[s_0(1+q_0)+j_0(6q_0^2+17q_0+3j_0+5) \nonumber \\&+&3q_0(5q_0^3+20q_0^2+29q_0+16)+9]+\frac{15}{8},
\end{eqnarray}
\begin{eqnarray}\label{f5T}
    \frac{f^{(v)}(T_0)}{(6H_0^2)^{-4}}&=&\frac{-3\Omega_{m0}}{32(1+q_0)^7}[l_0(1+q_0)^2+s_0(10q_0^3+43q_0^2+46q_0+13)+10j_0s_0(1+q_0)\nonumber \\&+&5j_0^2(6q_0^2+22q_0+3j_0+7)+j_0(45q_0^4+225q_0^3+412q_0^2+219q_0+32)\nonumber\\&+&3q_0(35q_0^5+210q_0^4+518q_0^3+666q_0^2
    +448q_0+150)+60]+\frac{105}{16},
\end{eqnarray}

\end{widetext}

\noindent Eqs.(\ref{fT}) - (\ref{f5T}) make it possible to estimate the present day values of $f(T)$ and its first five derivatives as function of the Hubble constant $H_0$ and the cosmographic parameters $(q_0, j_0, s_0, l_0)$ provided a value for the matter density parameter $\Omega_{m0}$ is given.

\subsection{The $\Lambda$CDM model}

In order to get a first hint on the possible values of $f(T)$ and its derivatives we have to reproduce
 the cosmographic parameters for $\Lambda$CDM model as  simplest case. This is a minimal approach but it is useful to probe the self-consistency of the model.
The cosmographic parameters for the $\Lambda$CDM model read

\begin{widetext}

\begin{equation}\label{qlambda}
    q=-(\frac{H_0}{H})^2(1-\Omega_{m0}-\frac{\Omega_{m0}}{2a^3}),
\end{equation}
\begin{equation}\label{jlambda}
    j=(\frac{H_0}{H})^3(1-\Omega_{m0}+\frac{\Omega_{m0}}{a^3})^{3/2},
\end{equation}
\begin{equation}\label{slambda}
    s=(\frac{H_0}{H})^4(1-2\Omega_{m0}-\frac{5\Omega_{m0}}{2a^3}+\Omega_{m0}^2+\frac{5\Omega_{m0}^2}{2a^3}-\frac{7\Omega_{m0}^2}{2a^6}),
\end{equation}
\begin{equation}\label{llambda}
    l=(\frac{H_0}{H})^5(1-2\Omega_{m0}+\frac{5\Omega_{m0}}{a^3}+\Omega_{m0}^2-\frac{5\Omega_{m0}^2}{a^3}+\frac{35\Omega_{m0}^2}{2a^6})(\sqrt{1-\Omega_{m0}
    +\frac{\Omega_{m0}}{a^3}}),
\end{equation}

\end{widetext}
which, evaluated at the present time, give
\begin{equation}\label{qzlambda}
    q_0=-1+\frac{3}{2}\Omega_{m0},
\end{equation}
\begin{equation}\label{jzlambda}
    j_0=1,
\end{equation}
\begin{equation}\label{szlambda}
    s_0=1-\frac{9}{2}\Omega_{m0},
\end{equation}
\begin{equation}\label{lzlambda}
    l_0=1+3\Omega_{m0}+\frac{27}{2}\Omega_{m0}^2.
\end{equation}

Inserting the previous equations in Eqs. (\ref{f2T}) - (\ref{f5T}), we obtain
\begin{equation}\label{constraint2}
    f''(T_0)=f'''(T_0)=f^{(iv)}(T_0)=f^{(v)}(T_0)=0,
\end{equation}
and in the absence of these terms $f(T)$ reduces to $f(T)\sim T-2\Lambda$. This is consistent with what we expected for the $\Lambda$CDM model and can be assumed as a consistency check.

\begin{figure*}
\centering
\subfigure{\includegraphics[width=5cm]{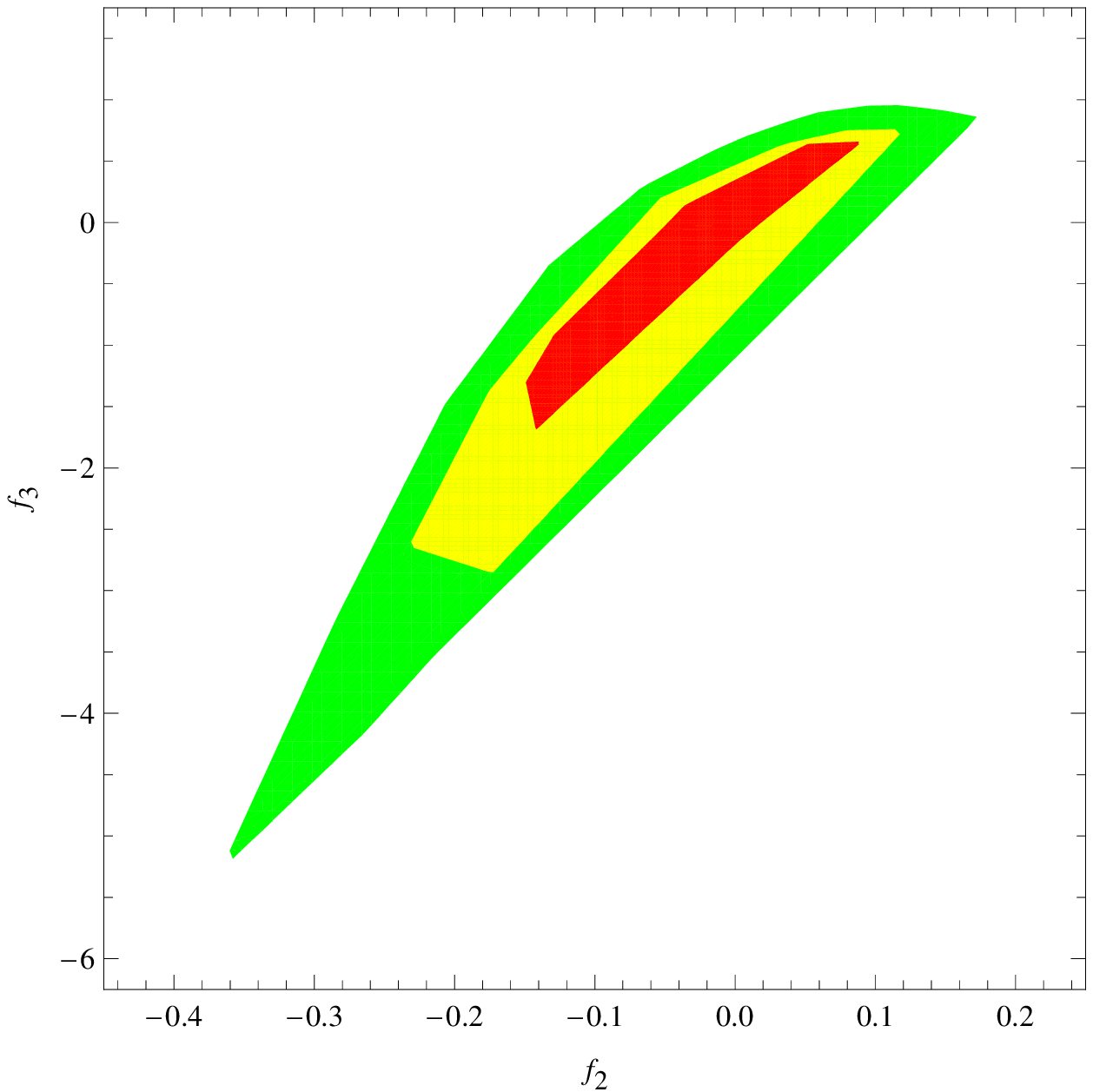}} \goodgap
\subfigure{\includegraphics[width=5cm]{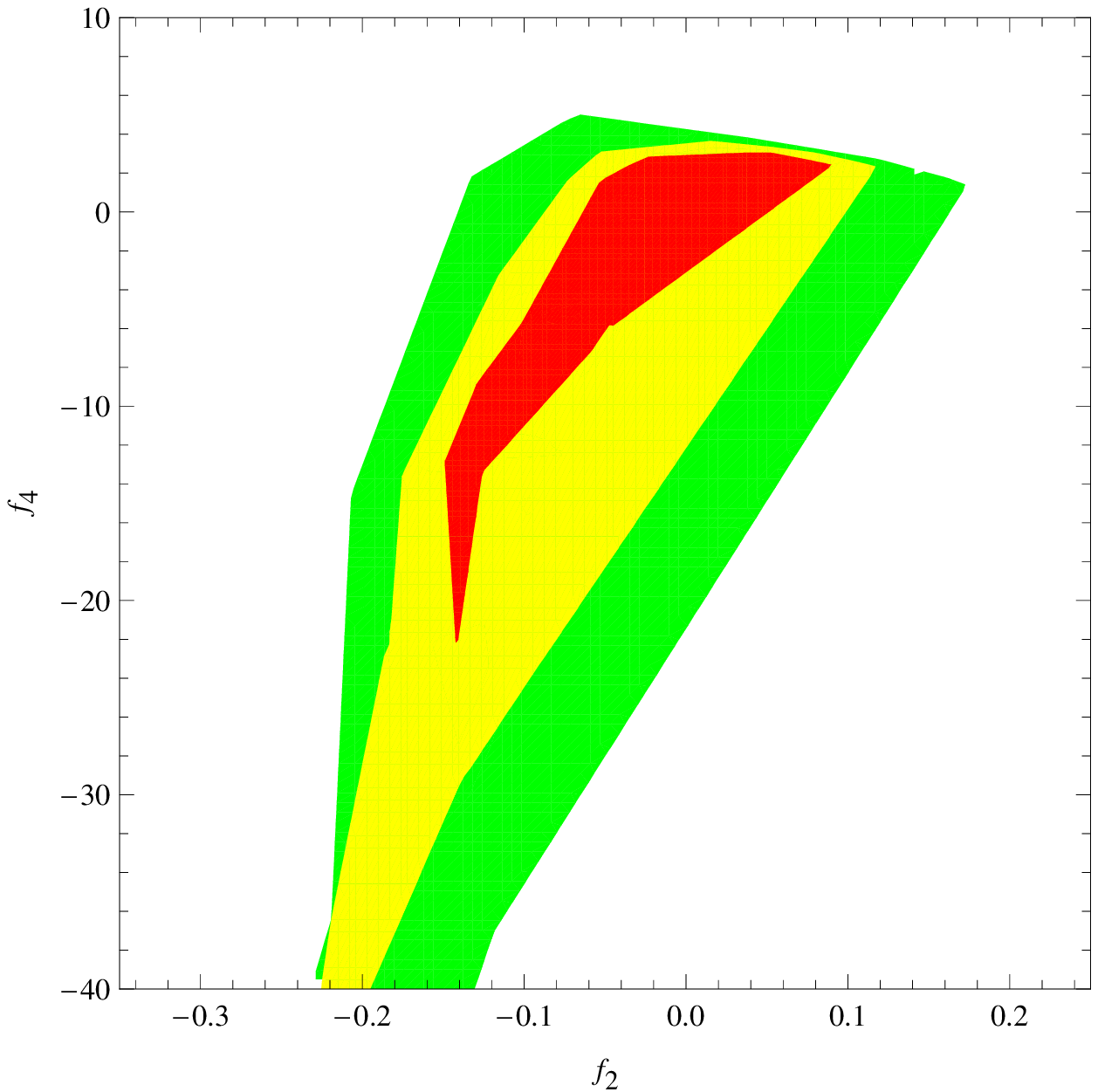}} \goodgap
\subfigure{\includegraphics[width=5cm]{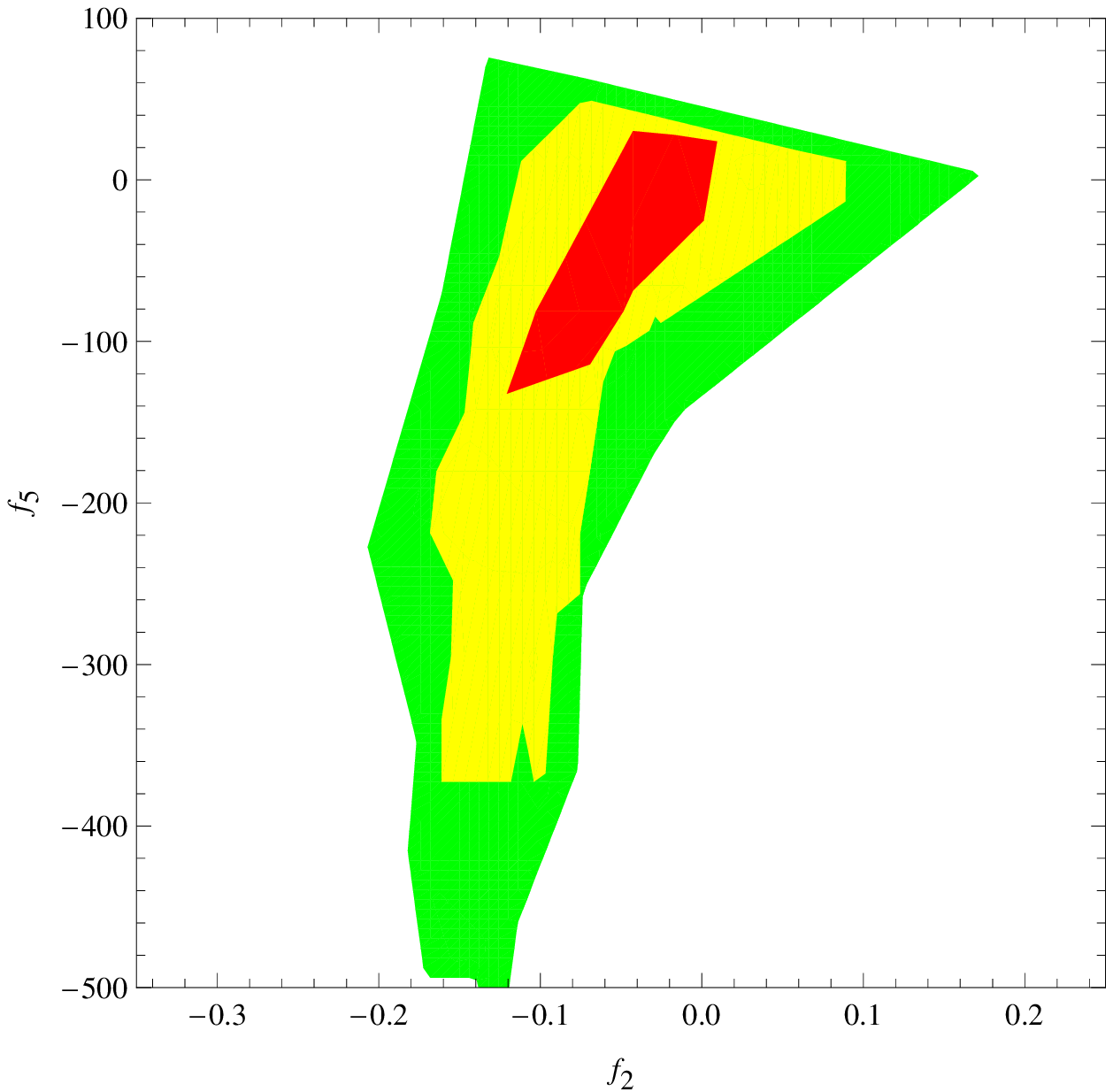}} \goodgap \\
\caption{Isolikelihood ($68$, $95$ and $99\%$ CL) contours for the $f_i$ quantities. The fuzzyness is due to numerical artifacts.}
\label{fig: clplots}
\end{figure*}

\section{Observational constraints}

In order to constrain the model by cosmography, i.e. to  estimate the function $f(T)$ through its own value and that of its derivatives at the present time, we need to constrain observationally the cosmographic parameters by using  appropriate distance indicators. Moreover, we must take care that the expansion of the distance related quantities in terms of $(q_0, j_0, s_0, l_0)$ closely follows the exact expressions over the range probed by the data used. Taking SNeIa and a fiducial $\Lambda$CDM model as a test case, one has to check that the approximated luminosity distance\footnote{See \cite{Capozziello:2008qc} for the analytical expression.} deviates from the $\Lambda$CDM one less than the measurement uncertainties up to $z \simeq  1.5$ to avoid introducing any systematic bias. Since we are interested in constraining $(q_0, j_0, s_0, l_0)$, we will expand the luminosity distance $D_L$ up to the fifth order in $z$ which indeed allows us to track the $\Lambda$CDM expression with an error less than $1\%$ over the full redshift range. We have checked that this is the case also for the angular diameter distance $D_A = D_L(z)/(1 + z)^2$ and the Hubble parameter $H(z)$ which, however, we expand only up to the fourth order to avoid introducing a further cosmographic parameter.

\begin{table}[t]
\begin{center}
\begin{tabular}{cccccc}
\hline
$x$ & $x_{BF}$ & $\langle x \rangle$ & $x_{med}$ & $68\%$ CL & $95\%$ CL \\
\hline \hline
~ & ~ & ~ & ~ & ~ & ~ \\
$h$ & 0.718 & 0.706 & 0.706 & (0.693, 0.719) & (0.679, 0.731) \\
~ & ~ & ~ & ~ & ~ & ~ \\
$q_{0}$ & -0.64 & -0.44 & -0.43 & (-0.60, -0.30) & (-0.71, -0.26) \\
~ & ~ & ~ & ~ & ~ & ~ \\
$j_{0}$ & 1.02 & -0.04 & -0.15 & (-0.88, -0.90) & (-1.07, 1.40) \\
~ & ~ & ~ & ~ & ~ & ~ \\
$s_{0}$ & -0.39 & 0.18 & 0.02 & (-0.57, 1.07) & (-1.04, 1.78) \\
~ & ~ & ~ & ~ & ~ & ~ \\
$l_{0}$ & 4.05 & 4.64 & 4.54 & (2.99, 6.48) & (1.78, 8.69) \\
~ & ~ & ~ & ~ & ~ & ~ \\
\hline
\end{tabular}
\end{center}
\caption{Constraints on the cosmographic parameters. Columns are as follows\,: 1. parameter id; 2. best fit; 3., 4. mean and median from the marginalized likelihood; 5., 6. $68$ and $95\%$ confidence ranges.}
\label{tab: cosmofit}
\end{table}

\begin{table}[t]
\begin{center}
\begin{tabular}{cccccc}
\hline
$x$ & $x_{BF}$ & $\langle x \rangle$ & $x_{med}$ & $68\%$ CL & $95\%$ CL \\
\hline \hline
~ & ~ & ~ & ~ & ~ & ~ \\
$f_{0}$ & -1.742 & -1.733 & -1.733 & (-1.743, -1.723) & (-1.751, -1.712) \\
~ & ~ & ~ & ~ & ~ & ~ \\
$f_{2}$ & -0.033 & 0.113 & 0.147 & (0.007, 0.208) & (-0.153, 0.226) \\
~ & ~ & ~ & ~ & ~ & ~ \\
$f_{3}$ & -0.092 & 0.530 & 0.815 & (0.172, 0.921) & (-1.483, 1.033) \\
~ & ~ & ~ & ~ & ~ & ~ \\
$f_{4}$ & 0.294 & -0.955 & 1.061 & (0.193, 2.306) & (-18.307, 3.603) \\
~ & ~ & ~ & ~ & ~ & ~ \\
$f_{5}$ & 8.690 & -68.893 & 6.371 & (2.956, 11.014) & (-370.966, 31.004) \\
~ & ~ & ~ & ~ & ~ & ~ \\
\hline
\end{tabular}
\end{center}
\caption{Constraints on the $f_i$ values from the Markov Chain for the cosmographic parameters. Columns order is the same as in Table\,\ref{tab: cosmofit}.}
\label{tab: ftfit}
\end{table}

In order to constrain the parameters $(h, q_0, j_0, s_0, l_0)$, Bouhmadi\,-\,Lopez et al. \cite{blcc10} have used the Union2 SNeIa dataset \cite{Union2} and the BAO data from the analysis of the SDSS seventh release \cite{P10} adding a prior on $h$ from the recent determination of the Hubble constant by the SHOES team \cite{shoes}. We update here their analysis adding the measurement of $H(z)$ obtained in \cite{Stern} from the age of passively evolving galaxies and in \cite{Cabre} from the radial BAO. Exploring the five dimensional parameter space with a Markov Chain Monte Carlo method, we obtained the constraints summarized in Table\,\ref{tab: cosmofit} in agreement with previous results in literature \cite{Vitagliano:2009et,XW10,Enzo2011}. Note that, because of the degeneracies among the five cosmographic parameters, the best fit values can also be different from the median ones, which is indeed what happens here. This is, however, not a shortcoming of the fitting analysis, but a consequence of the Bayesian approach giving more importance to sampling the marginalized parameters distributions rather than to looking for the best fit accordance within a given model and the available dataset\footnote{Qualitatively, one can say that the best fit value of, e.g., $q_0$ is less important than the median one since the best fit $q_0$ is the correct one only if the other parameters also take their best fit values, while the median one is more reliable since it describes the full distribution whatever are the values of the other parameters.}. In particular, here, the best fit values are quite close to those predicted for the $\Lambda$CDM model (for instance, $j_0 = 1$ for a $\Lambda$ dominated universe), while the median ones allow for significant deviations (with the $\Lambda$CDM values being, however, within the $95\%$ confidence ranges).

In order to translate the constraints on the cosmographic parameters on similar constraints on the present day values of $f(T)$ and its derivatives, we should just use Eqs.({\ref{fT})\,-\,(\ref{f5T}) evaluating them along the final coadded and thinned chain of the cosmographic parameters and then looking at the corresponding histograms. To this end, however, we should set the value of $\Omega_{m0}$ which is not constrained by the fitting analysis described before. To overcome this difficulty, we rely on the WMAP7 determination of the physical matter density $\omega_m = \Omega_{m0} h^2 = 0.1329$ and, for each value of $h$ along the chain, we fix $\Omega_{m0} = \omega_m/h^2$ having neglected the error on $\omega_m$ since it is subdominant with respect to the one on $h$. Note that the adopted estimate of $\omega_m$ comes from the fit to the CMBR anisotropy spectrum and mainly depends on the early universe physics only. Since it is reasonable to expect that GR is recovered in this limit, we can safely assume the validity of this result whichever $f(T)$ model is considered. Defining for shortness

\begin{displaymath}
f_n = f^{(n)}(T_0)/(6 H_0^2)^{-(n-1)} \ ,
\end{displaymath}
we finally get the constraints summarized in Table\,\ref{tab: ftfit} and shown in Fig.\,\ref{fig: clplots} where the degeneracy between some couples of parameters is shown as an example. Note that, as best fit value, we mean the one obtained by fixing the cosmographic parameters to the best fit values. However, because of the degeneracies among $(q_0, j_0, s_0, l_0)$ and the nonlinear behavior of the relations with $f_n$, it is possible that the best fit $f_n$ are quite different from their median values which is indeed the case (in particular, for $f_5$). Note also that the confidence ranges become larger as the order $n$ of the derivative increases. This is indeed an expected result since the higher is $n$, the larger is the number of cosmographic parameters involved so that the weakness of the constraints on the higher order cosmographic parameters and the degeneracies among them makes the constraints on $f_n$ weaker and weaker as $n$ gets larger. From a different point of view, such a behavior simply reflects the naive expectation that one has to go to deeper redshifts to probe the exact functional shape of $f(T)$ and hence put severe constraints on the value of its high order derivatives.   As a further remark, we note that the constraints on $(f_3, f_4, f_5)$ are strongly asymmetric with a long tail extending towards negative values causing a large offset between the mean and the median. This is actually a consequence of the term $(1 + q_0)^{-\alpha}$, with $\alpha = (3, 5, 7)$ for $(f_3, f_4, f_5)$ respectively, which enters as a common factor in Eqs.(\ref{f3T})\,-\,(\ref{f5T}). As $q_0$ comes close to -1, these term becomes increasingly large so thus making $f_{n}$ explode. It is, however, wort noting that values of $q_0$ close to -1 are indeed quite unlikely (although still allowed by our fit to a limited dataset) so that only the $95\%$ confidence ranges are affected.

\subsection{Dependence on the expansion order}

Although we have checked that our fifth order expansion closely matches the exact luminosity and angular diameter distances and the Hubble parameter within less than $1\%$, it is worth noting that a decent approximation is also obtained if we stop the expansion to the third or fourth order. Cutting the expansion to order three (four) means that we can only constrain cosmographic parameters up to the jerk $j_0$ (the snap $s_0$) and hence work out confidence ranges for the $f(T)$ derivatives up to the third (fourth) order. It is nevertheless worth exploring how the constraints depend on the order of the expansion. To this end, we fit the same dataset as above with both the third and fourth order expansion of the involved quantities and then use the corresponding Markov Chains to estimate confidence limits on $(f_0, f_2, f_3)$. From the third order fit, we get (median and $68$ and $95\%$ CL)\,:

\begin{displaymath}
\left \{
\begin{array}{l}
f_0 = -1.741_{-0.008 \ -0.016}^{+0.009 \ +0.017} \\
~ \\
f_2 = 0.005_{-0.069 \ -0.154}^{+0.054 \ +0.098} \\
~ \\
f_3 = 0.097_{-0.515 \ -1.552}^{+0.303 \ +0.475} \\
\end{array}
\right . \ ,
\end{displaymath}
while the fourth order fit gives\,:

\begin{displaymath}
\left \{
\begin{array}{l}
f_0 = -1.733_{-0.009 \ -0.019}^{+0.011 \ +0.021} \\
~ \\
f_2 = 0.043_{-0.076 \ -0.195}^{+0.061 \ +0.113} \\
~ \\
f_3 = 0.439_{-0.441 \ -1.536}^{+0.266 \ +0.439} \\
\end{array}
\right . \ .
\end{displaymath}
Comparing the value of $f_i$ for the different fits (including the fifth order one in Table\,\ref{tab: ftfit}) allows us to draw some interesting lessons. First, although the median values are different, the confidence ranges are well overlapped thus indicating that the order of the expansion should not have any statistically meaningful impact on the constraints. However further and accurate studies have to be performed in order to confirm this statement. On the other hand, increasing the order of the expansion shifts away from the $\Lambda$CDM one (i.e., $f_i = 0$ for $i > 1$). This is, actually, a subtle effect of the degeneracy among the cosmographic parameters. Indeed, increasing the order $n$ of the expansion adds further parameters to the fit thus allowing for much more combinations of the cosmographic parameters able to fit well the same data. As a consequence, the constraints on $q_0$ will become weaker allowing for models with $q_0$ closer to $-1$ and hence $(f_2, f_3)$ values far away from the fiducial $\Lambda$CDM ones. We, nevertheless, recommend the user to refer to the results in Table\,\ref{tab: ftfit} since the fifth order expansion provides a better approximation to the underlying expansion history so that the fit is less affected by any bias due to any error in the approximation.

\subsection{Deviations from basic assumptions}

The constraints discussed above have been obtained under two basic underlying assumptions. First, we have set $f^{\prime}(T_0) = 1$ in order to recover an effective gravitational constant  which matches the Newton one today. Actually, although reasonable, there are no compelling reasons why the Newton constant which is measured in laboratory experiments is the same as the cosmological one. As such, it is worth wondering how our results would change should we allow for deviations from the $G_N = G_{cosmo}$ assumption.

On the other hand, we have used the WMAP7 constraints on the physical matter density $\omega_M$ to infer the present day matter density parameter and then use Eqs.(\ref{fT})\,-\,(\ref{f5T}) to constrain $(f_0, f_2, f_3, f_4, f_5)$ from the cosmographic parameters. Some recent works \cite{pertft} have, however, investigate the evolution of perturbations in $f(T)$ theories finding out remarkable differences with respect to the standard GR. As a consequence, one can not exclude the possibility to recover a correct growth of structure even if $f(T))$ does not reduce to GR in the early universe. Should this be the case, the use of the WMAP7 $\omega_M$ value is incorrect.

Taking care of these possible effects is actually quite easy. Indeed, some algebra shows that Eqs.(\ref{fT})\,-\,\ref{f5T} can all be recast as\,:

\begin{displaymath}
f_n = \frac{{\cal{P}}_n(q_0, j_0, l_0, s_0)}{(1 + q_0)^{\alpha_n}} \Omega_M + \kappa_n (1 + \varepsilon)
\end{displaymath}
with ${\cal{P}}_n(q_0, j_0, l_0, s_0)$ a polynomial function of its arguments, $\alpha_n = (0, 1, 3, 5, 7)$ for $n = (0, 2, 3, 4, 5)$, $\kappa_n$ a constant depending on $n$ and we have set $f^{\prime}(T_0) = 1 + \varepsilon$. Using this simple formula allows to immediately scales our constraints to different values of the $\omega_M$ and $\varepsilon$ provided one has a theoretical or observational estimate of these quantities.

\section{Cosmography vs $f(T)$ models}

Up to now, we have never assumed any functional shape for $f(T)$ so that the constraints in Table I indeed holds for the full class of TEGR theories provided one can approximate $f(T)$ by its fifth order Taylor series over the redshift range probed by the data. Such a result can also be read in a different way. Given a $f(T)$ model, its characterizing parameters must be chosen in such a way that the constraints in Table I are satisfied. This consideration offers an interesting route to check the viability of a given $f(T)$ model without the need of explicitly solving the field equations and fitting the data.

As an example, let us assume the following model \cite{Myrza}\,:

\begin{equation}
f(T) = \alpha T + \beta T^{\delta} \ln{T} \ .
\label{eq: ftmyrza}
\end{equation}
Imposing Eq.(\ref{fT}) and $f'(T_0) = 1$ gives\,:

\begin{equation}
\alpha = \frac{2-\Omega_{m0} - [1 + (\Omega_{m0} - 2) \delta] \ln{T_0}}{1 + (\delta - 1) \ln{T_0}} \ ,
\label{eq: am}
\end{equation}

\begin{equation}
\beta = \frac{( \Omega_{m0}-1) T_0^{1 - \delta}}{1 + (\delta - 1) \ln{T_0}} \ ,
\label{eq: bm}
\end{equation}
so that we can express $f_i$ for $i = (2, 3, 4, 5)$ as function of $\delta$ only. We then proceed as follows. For each $f_2$ value of the sample obtained above from the cosmographic parameters analysis, we solve $\hat{f}_2(\delta) = f_2$. Since this equation has two roots, we store them and then compute $(f_3, f_4, f_5)$ for both values thus obtaining an histogram for the model prediction of these quantities. The median and $68\%$ and $95\%$ confidence ranges read\,:

\begin{displaymath}
\left \{
\begin{array}{l}
f_3 = -0.296_{-0.115 \ -0.149}^{+0.272 \ +0.599} \\
~ \\
f_4 = 0.891_{-0.799 \ -1.797}^{+0.330 \ +0.424}  \\
~ \\
f_5 = -3.568_{-1.274 \ -1.633}^{+3.143 \ +7.176} \\
\end{array}
\right . \ ,
\end{displaymath}
choosing the lowest $\delta$ solution and

\begin{displaymath}
\left \{
\begin{array}{l}
f_3 = 8.779_{-0.088 \ -0.131}^{+0.193 \ +0.415} \\
~ \\
f_4 = -3.120_{-0.024 \ -0.050}^{+0.018 \ +0.032}  \\
~ \\
f_5 = -31.033_{-0.811 \ -1.810}^{+0.371 \ +0.525} \\
\end{array}
\right . \ ,
\end{displaymath}
for the larger solution.  Since the $95\%$ CL in Table\,\ref{tab: ftfit} are quite large because of the impact of $q_0$, we will use only the $68\%$ confidence ranges which we compare the above constraints to. For the lower $\delta$ solutions, both $f_3$ and $f_5$ are smaller than the $68\%$ CL from cosmographic parameters, while the range for $f_4$ has a marginal overlap. On the other hand, choosing the largest $\delta$ solution leads to $(f_3, f_4, f_5)$ values that fully disagree with the model independent constraints. We therefore argue that the model (\ref{eq: ftmyrza}) is disfavored by the observational data.

In \cite{Myrza}, another model was also proposed\,:

\begin{equation}
f(T) = \alpha T + \beta T^n
\label{eq: ftmyrzabis}
\end{equation}
where, imposing as before the constraints on $f(T_0)$ and $f'(T_0)$, one easily gets\,:

\begin{equation}
\alpha = \frac{(2-\Omega_{m0} ) n - 1}{n - 1} \ ,
\label{eq: ambis}
\end{equation}

\begin{equation}
\beta = \frac{(\Omega_{m0}-1) T_0^{1 - n}}{n - 1} \ .
\label{eq: bmbis}
\end{equation}
We then solve $\hat{f}_2(n) = f_2$ and estimate the theoretically expected values for the other derivatives obtaining\,:

\begin{displaymath}
\left \{
\begin{array}{l}
f_3 = -0.285_{-0.116 \ -0.148}^{+0.272 \ +0.599} \\
~ \\
f_4 = 0.841_{-0.801 \ -1.802}^{+0.331 \ +0.424}  \\
~ \\
f_5 = -3.317_{-1.282 \ -1.641}^{+3.156 \ +7.215} \\
\end{array}
\right . \ .
\end{displaymath}
These values are still in disagreement with the constraints in Table\,\ref{tab: ftfit} hence making us argue against this model too. Actually, some caution is needed in this case. If we set $\alpha = 1$ and $|n|$ small enough, Eq.(\ref{eq: ftmyrzabis}) predicts an expansion rate which can be made arbitrarily close to the $\Lambda$CDM one. Indeed, if we use the best fit value of the cosmographic parameters, we find $n = -0.011$ and quite small values for $(f_3, f_4, f_5)$ as expected for $\Lambda$ term. Actually, the disagreement with the constraints in Table\,\ref{tab: ftfit} may be due to a failure of one of the underlying assumptions in the derivation of Eqs.(\ref{fT})\,-\,(\ref{f5T}). Indeed, these relations have been obtained by Taylor expanding $f(T)$ to the fifth order thus implicitly assuming that the higher order terms are subdominant. Depending on the value of $n$, however, this assumption can fail for the model (\ref{eq: ftmyrzabis}) so that the constraints on $f_n$ should not be considered reliable.

\section{Conclusions}

Cosmography offers a valid tool to investigate cosmic expansion in a model independent way. The constraints on the cosmographic parameters $(q_0, j_0, s_0, l_0)$ obtained by fitting to SNeIa Hubble diagram and BAO data are fully general relying on the only assumption that the universe is homogenous and isotropic on large scales. As such, any given cosmological model should predict $(q_0, j_0, s_0, l_0)$ values which are in agreement with these constraints. Such a premise makes it clear why studying the cosmography of a given theory can offer a valuable help to check its viability as an explanation of the observed cosmic speed up.

Motivated by these considerations, we have  discussed the cosmography of TEGR theories obtaining the expression of $f(T)$ and its derivatives as a function of the matter density parameter $\Omega_{m0}$ and the cosmographic parameters $(h, q_0, j_0, s_0, l_0)$. It is worth stressing that the relations thus found hold for all TEGR models provided they can be well approximated by their fifth order Taylor expansion, at least over the redshift range probed by the data used to constrain the cosmography. A key role has been played by the assumption $f^{\prime}(T_0) = 1$, meaning  that the effective gravitational constant  equals the Newton one at redshift $z = 0$.  Although this is a quite reasonable assumption, it nevertheless rely on the underlying identification of the cosmological $G_{cosmo}$ with the local one $G_N$. Should $f^{\prime}(T_0) \ne 1$, one could re-derive our results, but the price to pay is to lower the order of the expansion of one degree. Alternatively, one can leave $f_1 = f^{\prime}(T_0)$ as a free parameter and check which is the impact on $(f_2, f_3, f_4, f_5)$. Not surprisingly, if $f_1 \simeq 1 + \varepsilon$ with $|\varepsilon| << 1$, one could still use our relations to constrain $(f_2, f_3, f_4, f_5)$ making a systematic error which is by far smaller than the statistical uncertainties unless  unreasonable large values $|\varepsilon|$ $(> 0.1)$ are adopted.

The above relations allow to transform the constraints on $(\Omega_{m0}, h)$ and the cosmographic parameters $(q_0, j_0, s_0, l_0)$ into similar ones for the $(f_2, f_3, f_4, f_5)$ quantities. Coming out from a model independent approach as cosmography, these constraints have to be fulfilled by any TEGR model. As such, we can investigate {\it a priori} (i.e., without solving the field equation) of a given $f(T)$ theory by simply comparing the theoretically predicted 
$(f_2, f_3, f_4, f_5)$ with the observed ones. As an application, we have considered here two particular classes showing that, although they can in principle give rise to an accelerated expansion, they are both unable to predict the observationally motivated $(f_2, f_3, f_4, f_5)$ values so that they can be rejected. Expanding on this idea, one could also reverse the approach and build up a class of theories that fits the above constraints from the beginning and then investigate which is the expansion history at higher $z$ and the growth of structure.

However,  we have to stress again
that the method proposed here has some shortcomings:  $i)$ the truncation of the
Taylor expansion at some predefined order can be problematic due to the dropped
higher-order terms. Such terms could  reveal  important and so an arbitrary truncation could be dangerous; $ii)$ one would face difficulty when
trying to retain as high an order as possible, because the high-order terms are
increasingly weakly constrained by the limited data and could have large errors
(for example,  the constraints given in Table II and those in
Sec. VA could not give $f(T)$-curves which agree very well with each other). In this case, 
the second shortcoming is made worse by the first one. 

To be more specific, the new constraints on the $(h_0, q_0, j_0, s_0, l_0)$
parameters make sense, as well as their translation to the $f_i$
($i=0,2,\cdots,5$) parameters, considered the fact that the higher
$i$ is, the bigger and more uncertain is $f_i$.  This is a weakness of the method
used here, namely the higher-order terms in the Taylor expansion, which could be
important in the overall behavior of $f(T)$, are more difficult to predict
accurately. It is worth noticing  that smaller deviations of $(h_0, q_0, j_0, s_0, l_0)$ from the
corresponding $\Lambda$CDM values could cause very big deviations of $f_i$s from their
corresponding $\Lambda$CDM values which are identically zero.

Another comment is in order for the results in Sec. VA. The  fits have been performed with different orders of the Taylor expansion. 
We can see that the $f_0$ parameter perfectly agrees with
the value in Table II, but not $f_2$ or $f_3$. The reason is 
 that we have used a different number of parameters to fit the  same curve.   
 The assumption is that the order of expansion has no  statistically meaningful impact on the constraints, but such a statement  should be confirmed by further studies using more complete data sets.
 On the other hand, it is clear that  the increasing  order of expansion shifts away from the
$\Lambda$CDM fiducial values.

Although these drawbacks, the approach  is interesting 
and might be made more accurate as soon as  more data, especially those coming from
higher-redshift surveys, will be taken into account. 

As a final comment, it is worth noticing how the renewed interest in old dated cosmography has now opened the way to an alternative and yet powerful method to investigate, on the same ground, both dark energy models and modified gravity theories, such as $f(R)$ and TEGR models. After so many years, however, we are no more interested in finding only two numbers, namely $(h, q_0)$, but rather we now need a fifth order expansion, hence five quantities, to constrain not only the evolution of the universe, but also the underlying theory of gravity.
\section*{Acknowledgments}

VFC is  supported by the Italian Space Agency (ASI).


\begin{thebibliography}{99}

\bibitem{Riess} A. G. Riess, et al., Astron. J. 116, 1009 (1998); S. Perlmutter, et al., Astrophys. J. 517,
565 (1999).
\bibitem{Spergel} D. N. Spergel, et al., ApJS, 148, 175 (2003); D. N. Spergel, et al., ApJS, 170, 377S (2007).
\bibitem{Tegmark} M. Tegmark, et al., Phys. Rev. D. 69, 103501 (2004).
\bibitem{Eisenstein} D. J. Eisenstein, et al., Astrophys. J. 633, 560 (2005).
\bibitem{Carroll} S. M. Carroll, Living Rev. Rel. 4 (2001) 1; E. J. Copeland, M. Sami and S. Tsujikawa, Int. J. Mod. Phys. D. 15, 1753 (2006).
\bibitem{Yang} R. J. Yang and S. N. Zhang, Mon. Not. R. Astron. Soc. 407, 1835 (2010).
\bibitem{Feng}  B. Feng, X. L. Wang and  X. M. Zhang, Phys. Lett. B. 607, 35 (2005).
\bibitem{Caldwell}  R. R. Caldwell,  R. Dave and  R. J. Steinhardt, Phys. Rev. Lett. 80, 1582 (1998).
\bibitem{Caldwell2}  R. R. Caldwell, Phys. Lett. B. 545, 23 (2002).
\bibitem{Armendariz}  C. Armendariz-Picon,  V. Mukhanov and  P. J. Steinhardt, Phys. Rev. D. 63, 103510 (2001).
\bibitem{Padmanabhan} T. Padmanabhan, Phys. Rev. D. 66, 021301 (2002); A. Sen, Phys. Scripta. T. 117, 70 (2005).
\bibitem{eli}  E. Elizadle, S. Nojiri and  S. D. Odintsov, Phys. Rev. D. 70, 043539 (2004).
\bibitem{Kamenshchik}  A. Kamenshchik, U. Moschella and V. Pasquier, Phys. Lett. B. 511, 265 (2001).
\bibitem{Bento}  M. C. Bento, O. Bertolami and A. A. Sen, Phys. Rev. D. 66, 043507 (2002).
\bibitem{Cohen}  A. G. Cohen, D. B. Kaplan and A. E. Nelson, Phys. Rev. Lett. 82, 4971 (1999); M. Li, Phys. Lett. B. 603, 1 (2004).
\bibitem{Wei} H. Wei and R. G. Cai, Phys. Lett. B. 663, 1 (2008); H. Wei and R. G. Cai, Phys. Lett. B. 660, 113 (2008).
\bibitem{Gao} C. Gao, F. Wu, X. Chen and Y. G. Shen, Phys. Rev. D. 79, 043511 (2009).
\bibitem{capfra} S. Capozziello, M. Francaviglia , Gen.
Rel. Grav.  40, 357 (2008).
\bibitem{Nojiri}  S. Nojiri and S.D. Odintsov,  	arXiv:1011.0544  [gr-qc],  to appear in  Phys. Rep. (2011).
\bibitem{book} S. Capozziello  and  V. Faraoni, {\it Beyond Einstein Gravity}, Fundamental Theories of Physics Vol. 170, Springer Ed., Dordrecht  (2011).
\bibitem{sotfa}T.P. Sotiriou, V. Faraoni, Rev. Mod. Phys. 82, 451 (2010).
\bibitem{Mota} A. De Felice, D. F. Mota and S. Tsujikawa, Phys. Rev. D. 81, 023532 (2010).
\bibitem{Farajollahi} H. Farajollahi, M. Farhoudi and H. Shojaie, Int. J. Theor. Phy. 49, 10, 2558 (2010).
\bibitem{Zuntz} J. Zuntz, T. G. Zlosnik, F. Bourliot, P. G. Ferreira and G. D. Starkman,  Phys. Rev. D. 81, 104015 (2010).
\bibitem{Camera} M. La Camera, Mod. Phys. Lett. A. 25, 781-792 (2010).
\bibitem{noj1} S. Nojiri and S. D. Odintsov, Phys. Rev. D. 68, 123512 (2003).
\bibitem{noj2} S. Nojiri and S. D. Odintsov, Phys. Rev. D. 74, 086005 (2006).
\bibitem{noj3} S. Nojiri, and S.D. Odintsov, Int. J. Geom. Meth. Mod. Phys. 4, 115-146 (2007).
\bibitem{noj4} M. C. B. Abdalla, S. Nojiri and S. D. Odintsov,  Class. Quant. Grav. 22, L35 (2005).
\bibitem{noj5} S. Nojiri and S. D. Odintsov, Phys. Rev. D. 77, 026007 (2008).
\bibitem{Einstein} A. Einstein, Sitz. Preuss. Akad. Wiss. p. 217; ibid p. 224  (1928).
\bibitem{Einstein2} A. Einstein (2005), translations of Einstein papers by A. Unzicker and T. Case, (arXiv:physics/0503046).
\bibitem{Hoff} J. M. Hoff da Silva and R. da Rocha, Phys. Rev. D. 81, 024021 (2010).
\bibitem{Hayashi} K. Hayashi and T. Shirafuji, Phys. Rev. D. 19, 3524 (1979), Addendum-ibid. D. 24, 3312 (1982).
\bibitem{Ferraro3} R. Ferraro and F. Fiorini, Phys. Rev. D. 78, 124019 (2008).
\bibitem{Ulhoa} S. C. Ulhoa, J. F. da Rocha Neto and J. W. Maluf, Int. J. Mod. Phys. D. Vol. 19, No. 12, 1925-1935 (2010).
\bibitem{Nashed} G. G. L. Nashed, Int. J. Mod. Phys. A. Vol. 25, No. 14, 2883-2895 (2010).
\bibitem{Sharif} M. Sharif and S. Taj, Mod. Phys. Lett. A. 25, 221-232 (2010).
\bibitem{Lucas} T. G. Lucas, Y. N. Obukhov and J.G. Pereira, Phys. Rev. D. 80, 064043 (2009).
\bibitem{Ferraro2} R. Ferraro and F. Fiorini, Phys. Rev. D. 75, 084031 (2007).
\bibitem{Poplawski} N. J. Poplawski, Phys. Lett. B. 694, 181-185 (2010).
\bibitem{Wu} P. Wu and H. Yu, Eur. Phys. J. C. 71, 1552 (2011).
\bibitem{Wu1} P. Wu and H. Yu, Phys. Lett. B. 693, 415-420 (2010).
\bibitem{Ao} X. C. Ao, X.Z. Li and P. Xi, Phys. Lett. B. 694, 186-190 (2010).
\bibitem{Bengochea} G. R. Bengochea, Phys. Lett. B 695-405 (2011).
\bibitem{Ferraro} G. Bengochea and R. Ferraro, Phys. Rev. D. 79, 124019 (2009).
\bibitem{Wu2} P. Wu and H. Yu, Phys. Lett. B. 692, 176-179 (2010).
\bibitem{Yang2} R. J. Yang, (arXiv:1007.3571v2).
\bibitem{lorentz} B.Li, T. P. Sotiriou, J. D. Barrow,Phys. Rev. D83:064035 (2011).

\bibitem{Capozziello:2008qc}
  S.~Capozziello, V.~F.~Cardone and V.~Salzano,
  Phys.\ Rev.\  D  78 (2008) 063504

\bibitem{blcc10}
M. Bouhmadi\,-\,L\'opez, S. Capozziello, V.F. Cardone, Phys. Rev. D, 82, 103526

\bibitem{Stern}
D. Stern, R. Jimenez, L. Verde, S.A. Stanford, M. Kamionkowski, ApJS, 188, 280, 2010

\bibitem{Cabre}
E. Gaztanaga, A. Cabr\'e, L. Hui, MNRAS, 399, 1663, 2009

\bibitem{Union2}
R. Amanullah, C. Lidman, D. Rubin, G. Aldering, P. Astier et al., ApJ, 716, 712, 2010

\bibitem{P10}
W.J. Percival, B.A. Reid, D.J. Eisenstein, N.A. Bahcall, T. Budavari, et al., MNRAS, 401, 2148, 2010

\bibitem{shoes}
A.G. Riess, L. Macri, W. Li, H. Lampeitl, S. Casertano et al., ApJ 699, 539, 2009

\bibitem{Vitagliano:2009et}
  V.~Vitagliano, J.~Q.~Xia, S.~Liberati and M.~Viel,
  JCAP  1003 (2010) 005
  [arXiv:0911.1249 [astro-ph.CO]].

\bibitem{XW10}
L. Xu, Y. Wang, preprint arXiv\,:1009.0963, 2010

\bibitem{Enzo2011}
S. Capozziello, R. Lazkoz, V. Salzano, preprint arXiv\,:1104.3096, 2011

\bibitem{pertft}
J.B. Dent, S. Dutta, E.N. Saridakis, JCAP, 01, 009, 2011;
S.H. Chen, J.B. Dent, S. Dutta, E.N. Saridakis, Phys. Rev. D, 83, 023508, 2011
R. Zheng, Q.G. Huang, JCAP, 03, 002, 2011;
B. Li, T.P. Sotiriou, J.D. Barrow, Phys. Rev. D, 83, 104017, 2011

\bibitem{Myrza}
R. Myrzakulov, preprint arXiv\,:1006.1120, 2010

\end{thebibliography}
\end{document}